\documentstyle[aps,prd,multicol]{revtex}
\tighten

\begin{document}

\title{A new sequence of topological terms at any spacetime
dimensions}

\author{J. Barcelos-Neto$^a$ and E.C. Marino$^b$}

\address{\mbox{}\\
Instituto de F\'{\i}sica\\
Universidade Federal do Rio de Janeiro\\
RJ 21945-970 - Caixa Postal 68528 - Brasil}
\date{\today}

\maketitle
\begin{abstract}
\hfill{\small\bf Abstract\hspace*{1.7em}}\hfill\smallskip
\par
\noindent

We investigate a sequence of quadratic topological terms of the
Chern-Simons type in different spacetime dimensions, related by
dimensional compactification and sharing the properties of topological
mass generation and statistical transmutation. The implications for
bosonization
in several dimensions are also analyzed.

\end{abstract}

\pacs{PACS numbers: 11.10.Ef, 11.10.Kk, 11.15.-q}
\smallskip\mbox{}

\begin{multicols}{2}

{\bf 1.} The Chern-Simons term is usually considered at spacetime
dimension $D=2+1$ and is given by (up to a mass dimension constant)
\cite{Deser}

\begin{equation}
S_{CS}=\int d^3x\,\epsilon^{\mu\nu\rho}\,
\partial_\mu A_\nu A_\rho
\label{1.1}
\end{equation}

\noindent
We are going to refer to the Abelian case only. Considering that the
0+1 dimensional case is just given by $\int dt\,A$ \cite{Dunne}, we
may generalize the form of Chern-Simons term for any dimensions as

\begin{eqnarray}
S_{CS1}&=&\int dx\,A
\nonumber\\
S_{CS2}&=&\int d^2x\,\epsilon^{\mu\nu}\,\partial_\mu A_\nu
\nonumber\\
S_{CS3}&=&\int d^3x\,\epsilon^{\mu\nu\rho}\,
\partial_\mu A_\nu A_\rho
\nonumber\\
S_{CS4}&=&\int d^4x\,\epsilon^{\mu\nu\rho\lambda}\,
\partial_\mu A_\nu\partial_\rho A_\lambda
\nonumber\\
S_{CS5}&=&\int d^5x\,\epsilon^{\mu\nu\rho\lambda\eta}\,
\partial_\mu A_\nu\partial_\rho A_\lambda A_\eta
\nonumber\\
&\vdots&
\label{1.2}
\end{eqnarray}

The case of D=3, given by (\ref{1.1}), however, is special for many
reasons. Firstly, it is the only one that is quadratic in $A_\mu$
(note that in D=4 and actually in any even dimension it is a total
derivative and, therefore, classically trivial).
In this sense, D=3 is the only
case where it can be used alone as a ``kinetic'' term for $A_\mu$.
Secondly, it is well-known that in this case it produces a change in
the statistics of the particles that couple to $A_\mu$ \cite{Wilczek}.
Another feature of $S_{CS3}$ is that, if we consider it together with
a Maxwell term, it generates a mass for $A_\mu$, the so called
topological mass generation.

\medskip
In view of the special characteristics of $S_{CS3}$ and the fact that
the number of gauge fields is not the same for all $D$ we also
conclude that these terms cannot be linked each other by any kind of
dimensional compactification. It would be very interesting, on the
other hand, to be able to generate a whole sequence of topological
terms by dimensional reduction. For this purpose, we consider the
possibility of constructing topological terms involving gauge fields
of different ranks. These terms are not necessarily trivial for even
spacetime dimensions. In fact, for $D=4$, Cremer and J. Scherk
\cite{Cremer} introduced the following term,

\begin{equation}
S_4 =\int d^4x\,\epsilon^{\mu\nu\rho\lambda}\,
\partial_\mu A_\nu B_{\rho\lambda}
\label{1.3}
\end{equation}

\noindent
where $B_{\mu\nu}$ is the antisymmetric Kalb-Ramon field \cite{Kalb}
whose gauge transformation is expressed in terms of a vector parameter
$\xi_\mu$ as

\begin{equation}
\delta B_{\mu\nu}=\partial_\mu\xi_\nu-\partial_\nu\xi_\mu
\label{1.4}
\end{equation}

\noindent
This term can be associated to a mechanism of mass generation for the
gauge fields $A_\mu$ or $B_{\mu\nu}$ \cite{Cremer,Lahiri1}.
Indeed, considering (\ref{1.3}) together with Maxwell terms for
$A_\mu$ and $B_{\mu\nu}$, namely

\begin{equation}
S=\int d^4x\left(\frac{1}{12}\,H_{\mu\nu\rho}^2
-\frac{M}{2}\,\epsilon^{\mu\nu\rho\lambda}
A_\mu\partial_\nu B_{\rho\lambda}
-\frac{1}{4}\,F_{\mu\nu}^2\right)
\label{e1}
\end{equation}

\noindent
we get

\begin{equation}
S_{eff}[A_\mu]=-\frac{1}{4}\int d^4x\,F_{\mu\nu}
\left(1+\frac{M^2}{\Box}\right)F^{\mu\nu}
\label{e2}
\end{equation}

\noindent
or

\begin{equation}
S_{eff}[B_{\mu\nu}]=\frac{1}{12}\int d^4x\,H_{\mu\nu\rho}
\left(1+\frac{M^2}{\Box}\right)H^{\mu\nu\rho}
\label{e3}
\end{equation}

\noindent
respectively, upon integration over $B_{\mu\nu}$ or $A_\mu$. It is
opportune to mention that its non-Abelian version \cite {Lahiri2} can
also be used as an alternative mechanism of mass generation for the
gauge fields in the electroweak theory without Higgs bosons
\cite{Barc1}.

\medskip
The topological term given by (\ref{1.3}) presents the three features
closely related to the ones of (\ref{1.1}). It is quadratic in the
fields, it generates a mass for these fields and it is also related to
the change of statistics of extended objects, namely, strings, in $D=
3+1$ \cite{st}. It is, therefore, natural, attempting to find another
sequence of terms containing (\ref{1.1}) and (\ref{1.3}), and
presenting the common feature of being quadratic. We are going to see
that some other features are also shared.

\vspace{1cm}
{\bf2.} The motivation of the present paper is to make a general study
of a new sequence for topological terms, by considering that these
terms always have two fields (with the same rank or not). Since even
dimensional terms are not necessarily zero, our guide will be that
a term from a dimension $D$ can be obtained from compactification of a
term higher than $D$. The only natural constraint we impose is that
the usual Chern-Simons term in $D=2+1$ takes part in the sequence.

\medskip
In order to have a better comprehension of the problem, let us start
from a specific spacetime dimension, say $D=5$. At first sight, the
corresponding Chern-Simons term could be written by means of two gauge
fields of rank two, i.e. $\int d^5x\,\epsilon^{\mu\nu\rho\lambda\eta}
\,\partial_\mu B_{\nu\rho}B_{\lambda\eta}$. However, this term
corresponds to a total derivative and consequently is not a good
candidate for a Chern-Simons at spacetime dimension $D=5$. The next
one to play this role is formed by a totally antisymmetric gauge field
of rank three and a vector field, namely

\begin{equation}
S_5=\int d^5x\,\epsilon^{\mu\nu\rho\lambda\eta}\,
\partial_\mu A_\nu C_{\rho\lambda\eta}
\label{2.1}
\end{equation}

\noindent
The gauge transformation for the rank three gauge field
$C_{\mu\nu\rho}$ should be

\begin{equation}
\delta C_{\mu\nu\rho}=\partial_\mu\zeta_{\nu\rho}
+\partial_\rho\zeta_{\mu\nu}+\partial_\nu\zeta_{\rho\mu}
\label{2.2}
\end{equation}

\noindent
where $\zeta_{\mu\nu}$ is an antisymmetric gauge parameter.

\medskip
The consistency of the term given by (\ref{2.1}) can be verified by
performing the compactification of one space dimension and see if the
Chern-Simons term at $D=4$, given by (\ref{1.3}), is obtained. We use
the spontaneous compactification procedure \cite{Castellani}. Let us
compactify the coordinate $x^4$. Using a more convenient notation,
where capital Roman indices correspond to the spacetime dimension $D=
5$, we have

\begin{eqnarray}
&&\int d^5x\,\epsilon^{MNPQR}\,\partial_MA_NC_{PQR}
\nonumber\\
&&\phantom{\int d^5x\,}
=\epsilon^{\mu\nu\rho\lambda}\int d^4x\int_0^Rdx^4\,
\Bigl(\partial_4A_\mu C_{\nu\rho\lambda}
\nonumber\\
&&\phantom{\int d^5x\,=}
+\partial_\lambda A_4 C_{\mu\nu\rho}
+3\,\partial_\mu A_\mu C_{\rho\lambda4}\Bigr)
\label{2.3}
\end{eqnarray}

\noindent
where $\epsilon^{\mu\nu\rho\lambda}=\epsilon^{4\mu\nu\rho\lambda}$. We
next take the (Fourier) expansions:

\begin{eqnarray}
&&A_\mu(x,x^4)=\frac{1}{\sqrt R}\sum_{n=-\infty}^{\infty}
A_{(n)\mu}(x)\,\exp \Bigl(2in\pi\frac{x^4}{R}\Bigr)
\nonumber\\
&&A_4(x,x^4)=\frac{1}{\sqrt R}\sum_{n=-\infty}^{\infty}
\phi_{(n)}(x)\,\exp \Bigl(2in\pi\frac{x^4}{R}\Bigr)
\nonumber\\
&&C_{\mu\nu\rho}(x,x^4)=\frac{1}{\sqrt R}\sum_{n=-\infty}^{\infty}
C_{(n)\mu\nu\rho}(x)\,\exp \Bigl(2in\pi\frac{x^4}{R}\Bigr)
\nonumber\\
&&C_{\mu\nu4}(x,x^4)=\frac{1}{\sqrt R}\sum_{n=-\infty}^{\infty}
\tilde B_{(n)\mu\nu}(x)\,\exp \Bigl(2in\pi\frac{x^4}{R}\Bigr)
\label{2.4}
\end{eqnarray}

\noindent
Since all fields in (\ref{2.3}) are assumed to be real, we must have
$A^\ast_{(n)\mu}=A_{(-n)\mu}$, $\phi^\ast_{(n)}=\phi_{(-n)}$, and so
on.

\medskip
Considering similar expansions for the parameters related to the
gauge transformations of $A_M$ and $C_{MNP}$

\begin{eqnarray}
&&\alpha(x,x^4)=\frac{1}{\sqrt R}\sum_{n=-\infty}^{\infty}
\alpha_{(n)}(x)\,\exp \Bigl(2in\pi\frac{x^4}{R}\Bigr)
\nonumber\\
&&\zeta_{MN}(x,x^4)=\frac{1}{\sqrt R}\sum_{n=-\infty}^{\infty}
\zeta_{(n)MN}(x)\,\exp \Bigl(2in\pi\frac{x^4}{R}\Bigr)
\label{2.5}
\end{eqnarray}

\noindent
we obtain the following gauge transformations for the mode expansions
that appear in (\ref{2.4})

\begin{eqnarray}
&&\delta A_{(n)\mu}(x)=\partial_\mu\alpha_{(n)}(x)
\label{2.6}\\
&&\delta\phi_{(n)}(x)=\frac{2in\pi}{R}\,\alpha_{(n)}(x)
\label{2.7}\\
&&\delta C_{(n)\mu\nu\rho}(x)=\partial_\mu\zeta_{(n)\nu\rho}(x)
+\partial_\rho\zeta_{(n)\mu\nu}(x)
\nonumber\\
&&\phantom
{\delta C_{(n)\mu\nu\rho}(x)=\partial_\mu\zeta_{(n)\nu\rho}(x)}
+\partial_\nu\zeta_{(n)\rho\mu}(x)
\label{2.8}\\
&&\delta\tilde B_{(n)\mu\nu}(x)=\partial_\mu\tilde\xi_{(n)\nu}(x)
-\partial_\nu\tilde\xi_{(n)\mu}(x)
\nonumber\\
&&\phantom
{\delta C_{(n)\mu\nu\rho}(x)=\partial_\mu\zeta_{(n)\nu\rho}(x)}
+\frac{2in\pi}{R}\,\zeta_{(n)\mu\nu}(x)
\label{2.9}
\end{eqnarray}

\noindent
where

\begin{equation}
\tilde\xi_{(n)\mu}(x)=\zeta_{(n)\mu4}(x)
\label{2.10}
\end{equation}

Comparing (\ref{1.4}) and (\ref{2.9}), we observe that just $\tilde
B_{(0)\mu\nu}
$ can be identified with $B_{\mu\nu}$ where one takes $\tilde\xi_{(0)
\mu}=
\xi_\mu$.

\medskip
Let us now insert the expansions given by (\ref{2.4}) into
(\ref{2.3}). The final result is

\begin{eqnarray}
&&\int d^5x\,\epsilon^{MNPQR}\,\partial_MA_NC_{PQR}
\nonumber\\
&&\phantom{\int d^5x\,}
=\epsilon^{\mu\nu\rho\lambda}\sum_{n=-\infty}^\infty\int d^4x
\Bigl(\frac{2in\pi}{R}\,A_{(n)\mu}C^\ast_{(n)\nu\rho\lambda}
\nonumber\\
&&\phantom{\int d^5x\,=}
+\partial_\lambda\phi_{(n)}C^\ast_{(n)\mu\nu\rho}
+3\partial_\mu A_{(n)\nu}\tilde B^\ast_{(n)\rho\lambda}\Bigr)
\label{2.11}
\end{eqnarray}

\noindent
Since $\tilde B_{(0)\mu\nu}$ can be identified with the tensor gauge
field $B_{\mu\nu}$ and there is no problem in identifying $A_{(0)\mu}$
with $A_\mu$, the Cremer and Scherk topological term given by
(\ref{1.3}) is actually present in the compactified expression
(\ref{2.11}) for $n=0$. Further, there is another kind of topological
term that can also be identified in (\ref{2.11}) for $n=0$, involving
a real scalar field and a three-form gauge field, namely

\begin{equation}
S_4^\prime=\int d^4x\,\epsilon^{\mu\nu\rho\lambda}\,
\partial_\mu\phi\,C_{\nu\rho\lambda\eta}
\label{2.12}
\end{equation}

\noindent
where $\phi=\phi_{(0)}$. In fact, we could also have written a term
like this in $D=5$, involving a scalar and a four-form gauge field,

\begin{equation}
S_{5}^\prime=\int d^5x\,\epsilon^{MNPQR}\,\partial_M\phi\,D_{NPQR}
\label{2.13}
\end{equation}

\noindent
It is not difficult to see that the spontaneous compactification of
this term also leads to $S^\prime_4$ given by (\ref{2.12}), but not to
(\ref{1.3}).

\vspace{1cm}
{\bf 3.} We now proceed in a similar way and go to lower dimensions.
Having the same care in identifying vector, tensor and scalar fields
for $n=0$, we get from $D=4$ to $D=3$ the usual Chern-Simons terms
given by (\ref{1.1}) and also another one involving a scalar field,

\begin{equation}
S_3^\prime=\int d^3x\,\epsilon^{\mu\nu\rho}\partial_\mu\phi\,
B_{\mu\nu}
\label{3.1}
\end{equation}

\noindent
This term can be reached by compactification of both $S_4$ and
$S^\prime_4$, and has also been considered in a recent literature
\cite{Medeiros}. It is easily seen that topological terms involving
scalar field is a kind of residual Chern-Simons term that appears at
any spacetime dimensions. However, in $D=2$ and $D=1$ these terms are
the only ones that remain. Indeed,

\begin{eqnarray}
&&S_2=\int d^2x\,\epsilon^{\mu\nu}\partial_\mu A_\nu\,\phi
\label{3.2}\\
&&S_1=\int dx\,\dot\phi\,\varphi
\label{3.3}
\end{eqnarray}

\noindent
In the obtainment of (\ref{3.3}), the two different scalar fields come
from $A_{(0)1}$ and $\phi_{(0)}$.

\medskip
The term (\ref{3.1}) also has been shown to produce a mass generation
for the corresponding fields \cite{Medeiros}. For the one given by
(\ref{3.2}) and considering the kinetic terms for $\phi$ and $A_\nu$,
we have

\begin{equation}
S=\int d^2x\left[\frac{1}{2}(\partial_\mu \phi)^2
-M\epsilon^{\mu\nu}A_\mu\partial_\nu\phi
-\frac{1}{4} F_{\mu\nu}^2\right]
\label{e4}
\end{equation}

\noindent
After integrating over $\phi$ and $A_\nu$, we respectively obtain

\begin{equation}
S_{eff}[A_\mu]=-\frac{1}{4}\int d^2x\, F_{\mu\nu}
\left(1+\frac{M^2}{\Box}\right)F^{\mu\nu}
\label{e5}
\end{equation}

\noindent
and

\begin{equation}
S_{eff}[\phi]=\int d^2x\left[\frac{1}{2}(\partial_\mu \phi)^2
-\frac{M^2}{2}\phi^2\right]
\label{e6}
\end{equation}

\noindent
and we, once again, identify the mass generation mechanism. A similar
procedure also happens to (\ref{3.3}).

\medskip
We observe that the topological term in $D=1$ does not match the
corresponding one of the sequence given by (\ref{1.2}). Even though
that term appears naturally in the framework of quantum field theory,
in the sense that it can be generated by quantum corrections of
fermionic loops \cite{Dunne}, the $S_{CS1}$ term is not related to
mass generation. This property on the other hand is fulfilled by the
term given by (\ref{3.3}). Concerning the quantum generation for the
terms of the sequence we are studying, we remark that they are also
consistent from the quantum point of view. What happens is that they
emerge in an almost trivial way. For example, for the term given by
(\ref{1.3}), the $\epsilon^{\mu\nu\rho\lambda}$ tensor appears
directly from the interaction vertices involving fermion and tensor
fields, while the usual Chern-Simons term in $D=3$ comes from the
trace of gamma matrices in a one-loop calculation.

\vspace{1cm}
{\bf4.} In the previous section we have considered topological terms
by making spontaneous compactification starting from $D=5$. From the
results we have by now, it is feasible to infer those terms for
spacetime dimensions higher than five. For example, at $D=6$ these
terms are

\begin{eqnarray}
S_6&=&\int d^6x\,\epsilon^{\mu\nu\rho\lambda\eta\xi}\,
\partial_\mu\phi\,E_{\nu\rho\lambda\eta\xi}
\nonumber\\
S_6^\prime&=&\int d^6x\,\epsilon^{\mu\nu\rho\lambda\eta\xi}\,
\partial_\mu A_\nu\,D_{\rho\lambda\eta\xi}
\nonumber\\
S_6^{\prime\prime}&=&\int d^6x\,\epsilon^{\mu\nu\rho\lambda\eta\xi}\,
\partial_\mu B_{\nu\rho}\,C_{\lambda\eta\xi}
\label{4.1}
\end{eqnarray}

\noindent
Let us also write down the terms for $D=7$

\begin{eqnarray}
S_7&=&\int d^7x\,\epsilon^{\mu\nu\rho\lambda\eta\xi\zeta}\,
\partial_\mu\phi\,F_{\nu\rho\lambda\eta\xi\zeta}
\nonumber\\
S_7^\prime&=&\int d^7x\,\epsilon^{\mu\nu\rho\lambda\eta\xi\zeta}\,
\partial_\mu A_\nu\,E_{\rho\lambda\eta\xi\zeta}
\nonumber\\
S_7^{\prime\prime}&=&\int d^7x\,
\epsilon^{\mu\nu\rho\lambda\eta\xi\zeta}\,
\partial_\mu B_{\nu\rho}\,D_{\lambda\eta\xi\zeta}
\nonumber\\
S_7^{\prime\prime\prime}&=&\int d^7x\,
\epsilon^{\mu\nu\rho\lambda\eta\xi\zeta}\,
\partial_\mu C_{\nu\rho\lambda}\,C_{\eta\xi\zeta}
\label{4.2}
\end{eqnarray}

\noindent
and it is not difficult to infer the general case. We observe that the
number of terms increases with the spacetime dimension. There is just
one term for $D=1$ and $D=2$. For $D= 3,4,5$ there are two terms. In
fact, there would be three terms for $D=5$, but one of them, involving
two fields of rank two, is a total derivative. The same does not occur
for example for $D=7$, where the corresponding term involving two
gauge fields of rank three is not a total derivative. This just occurs
when the two equal gauge fields have even rank.

\medskip
The number of topological terms for a specific dimension $D$ is $D/2$
for $D$ even. In the case of odd $D$ we have that the number is
$D/2 - 1/2$ if the term with two equal gauge fields is a
total derivative and $D/2 + 1/2$ when it is not.

\medskip
An important point to be emphasized is that even though the number of
topological terms increases with $D$, there is one term at each
specific spacetime dimension that is more important than the others in
the sense that it generates all the other terms for lower dimensions.
It is not difficult to identity these terms. For example, at $D=5$
this term is $S_5$, in $D=6$, $S^{\prime\prime}_6$, and in $D=7$,
$S_7^{\prime\prime}$ (notice that it is not $S_7^{\prime\prime\prime}$
because it would not generate $S_6^{\prime}$).

\medskip
In what follows, we analyze physical properties of some of the
topological terms in the sequence, as far as statistical transmutation
and bosonization are concerned.

\vspace{1cm}
{\bf5.} Some of the topological terms generated in the sequence
studied in this paper play an important role in connection to the
statistical transmutation of different objects. In the case of $D=3$,
let us consider the coupling of a point-particle, associated to a
current density $j^\mu$, to the topological field:

\begin{equation}
{\cal L}_3 = \frac{\theta}{2} \epsilon^{\mu\alpha\beta}\,
A_\mu\partial_\alpha A_\beta-j^\mu A_\mu
\label{5.1}
\end{equation}

\noindent
In this equation,

\begin{equation}
j^\mu = \int_L d \xi^\mu \delta^3 (x - \xi)
\label{5.2}
\end{equation}

\noindent
where $L$ is the universe-line of the particle. The 0-component of the
field equation associated to expression (\ref{5.2}), for a static
point-particle is

\begin{equation}
j^0 =\delta^2 (\vec x - \vec x_0) = \theta \epsilon^{ij}
\partial_i A_j = \theta\,B
\label{5.3}
\end{equation}

\noindent
where $B = \epsilon^{ij}\partial_i A_j$ is the magnetic field, which
is a scalar in D=3. We see that the topological term imparts a point
magnetic flux to any point particle that couples to the $A_\mu$ field.
This fact is responsible for the change in the statistics of the
particle since, through an Aharonov-Bohm type effect, its wave
function will acquire a phase when interchanged with another
particle\cite{statt}.

\medskip
Let us consider now the case of a string in $D=4$. This is associated
to the $2$-tensor current density given by

\begin{equation}
j^{\mu\nu} = \int_S d^2 \sigma^{\mu\nu} \delta^4 (x - \xi)
\label{5.4}
\end{equation}

\noindent
where $S$ is the universe-sheet of the string. The coupling of the
string with the topological term $S_4$ is given by

\begin{equation}
{\cal L}_4 = \theta \epsilon^{\mu\nu\alpha\beta}\,
A_\mu \partial_\nu B_{\alpha\beta} - j^{\mu\nu} B_{\mu\nu}
\label{5.5}
\end{equation}

The $(0i)$-component of the field equation associated with the
Lagrangian density (\ref{5.5}), corresponding to a static string along
the spatial curve $\Gamma$, is

\begin{equation}
j^{0i} = \int_\Gamma d \xi^i \delta^3 (x - \xi)
= \theta\,B^i
\label{5.6}
\end{equation}

\noindent
where $B^i = \epsilon^{ijk}\partial_j A_k$ is the magnetic field. For
a straight string along the $x^3$-direction, piercing the
$(12)$-plane at $\vec x_0$, for instance, we have $\theta\,B^3 =
\delta^2 (\vec x - \vec x_0)$.

\medskip
We see that the topological term produces a constant magnetic field
along the string. It is not difficult to infer that charged strings in
the presence of the topological term $S_4$ will suffer a statistical
transmutation determined by $\theta$, again as a consequence of an
Aharonov-Bohm like effect. This fact has been already identified
before \cite{st} in a different framework. Nevertheless, it becomes
especially transparent here and can be unified with what happens in
other dimensions as well.

\medskip
Consider now the case of membranes in $D=5$. The current density
associated to a membrane is the $3$-tensor

\begin{equation}
j^{\mu\nu\alpha} = \int_V d^3 \sigma^{\mu\nu\alpha} \delta^5 (x - \xi)
\label{5.7}
\end{equation}

\noindent
where $V$ is the universe-volume of the membrane. The membrane couples
to the topological term $S_5$, Eq. (\ref{2.1}), in the following way:

\begin{equation}
{\cal L}_5 = \theta \epsilon^{\mu\nu\alpha\beta\gamma}\,
A_\mu \partial_\nu C_{\alpha\beta\gamma} - j^{\mu\nu\alpha}
C_{\mu\nu\alpha}
\label{5.8}
\end{equation}

The $(0ij)$-component of the field equation associated to (\ref{5.8})
and corresponding to a static membrane along the spatial surface
$\Sigma$ is

\begin{equation}
j^{0ij} = \int_\Sigma d^2 \xi^{ij} \delta^4 (x - \xi)
= \theta\,B^{ij}
\label{5.9}
\end{equation}

\noindent
where $B^{ij} = \epsilon^{ijkl}\partial_k A_l$ is the magnetic field,
which in $D=5$ is a $2$-tensor. We see that the topological term $S_5$
attaches a constant magnetic field along a membrane that couples to it
through the vector field $A_\mu$ in $D=5$. We may immediately conclude
that, in analogy to what happens in $D=3$ and $D=4$, a charged
membrane in the presence of the topological term $S_5$ will undergo
statistical transmutation determined by the parameter $\theta$. This
fact can certainly be also inferred through a study of the membrane
propagator or by the membrane creation operator formalism, as has been
done for the string in $D=4$ \cite{st}. We are presently investigating
this point.

\vspace{1cm}
{\bf6.} Another important feature associated with the sequence of
topological terms in an arbitrary dimension is bosonization. In this
case the basic starting point is $D=2$. Consider a point particle
associated to a current density $j^\mu$, which couples to the
topological term $S_2$, given by (\ref{3.2}), as follows

\begin{equation}
{\cal L}_2 = \theta \epsilon^{\mu\nu}\,
A_\mu \partial_\nu \phi - j^\mu A_\mu
\label{6.1}
\end{equation}

The $0$-component of the field equation obtained by varying
(\ref{6.1}) with respect to $A_\mu$, for a static point particle in
position $x_0$ is

\begin{equation}
j^0 = \delta (x - x_0) = \theta \partial_x \phi
\label{6.2}
\end{equation}

\noindent
This implies that corresponding to the point particle described by the
current $j^\mu$, we have a solution for the field $\phi$ given by

\begin{equation}
\phi_0 (x) = \theta\,\tilde\theta (x - x_0)
\label{6.3}
\end{equation}

\noindent
where $\tilde\theta (x - x_0)$ is the step function. This is a soliton
profile and we are led to infer that corresponding to the particles
associated with the current $j^\mu$, we will have soliton excitations
of the $\phi$-field, associated to the identically conserved current
$J^\mu = \epsilon^{\mu\nu} \partial_\nu \phi$. This fact is at the
very basis of the process of bosonization, whenever $j^\mu$ is a
fermionic current. The fermionic particles are identified with the
soliton excitations of the associated bosonic theory \cite{ems}.

\medskip
The previous observation allows us to look at the possibility of
bosonization in dimensions higher than $D=2$ and its possible
connection with the topological terms studied here. Indeed, from
(\ref{5.3}) we see that corresponding to a point particle associated
to a current $j^\mu$, minimally coupled to a Chern-Simons field in $D=
3$, we will have a magnetic vortex solution for the $A_\mu$-field at
the same point. These vortex configurations, however, are topological
solitons corresponding to the identically conserved current $J^\mu =
\epsilon^{\mu\nu\alpha} \partial_\nu A_\alpha$. Once more we see that
bosonization may be achieved by identifying the fermions associated to
a current $j^\mu$ with magnetic vortices in the corresponding bosonic
vector gauge field theory. This has actually been pursued in the case
of a free massless Dirac fermion \cite{em} but still there is a vast
area to explore in bosonization in $D=3$.

\medskip
Some conclusions can also be drawn by extending the present analysis
for $D=4$. Coupling a fermionic vector current to the vector field of
the topological term $S_4$, namely

\begin{equation}
{\cal L'}_4 = \theta \epsilon^{\mu\nu\alpha\beta}\,
A_\mu \partial_\nu B_{\alpha\beta} - j^{\mu} A_{\mu}
\label{6.4}
\end{equation}

\noindent
We see that the field equation obtained by varying with respect to
$A_{\mu}$ is

\begin{equation}
j^{\mu} = \theta \epsilon^{\mu\nu\alpha\beta}\,
\partial_\nu B_{\alpha\beta}
\label{6.5}
\end{equation}

\noindent
For a static point particle at $\vec x_0$, we have

\begin{equation}
j^{0} =\delta^3 (\vec x - \vec x_0) =
\theta \epsilon^{ijk}\,
\partial_i B_{jk}
\label{6.6}
\end{equation}

We see that the point fermionic particle may be identified with a
configuration of the bosonic Kalb-Ramond field $B_{\alpha\beta}$ with
a nonzero topological charge corresponding to the identically
conserved current $J^\mu = \epsilon^{\mu\nu\alpha\beta}\, \partial_\nu
B_{\alpha\beta}$. This observation should be at the basis of any
attempt to extend bosonization to $D=4$.

\vspace{1cm}
\noindent
{\bf Acknowledgment:}
This work is supported in part by CNPq-PRONEX 66.2002/1998-9 and also
by FAPERJ. J.B.-N. and E.C.M. are partially supported by CNPq.

\end{multicols}
\end{document}